\documentclass{jpp}
\usepackage{graphicx}
\usepackage{hyperref}
\usepackage{amsmath,amsfonts,amssymb}
\usepackage{csquotes}
\usepackage[ruled,vlined,resetcount]{algorithm2e}
\SetAlCapNameFnt{\footnotesize}
\SetAlCapFnt{\footnotesize}
\SetAlFnt{\footnotesize}

\usepackage{changes}

\newcommand{\muphy}{\texttt{muphyII}\xspace}

\shorttitle{Asymptotic-Preserving Two-Species and MHD Coupling}
\shortauthor{M. Deisenhofer, A. Mustonen, S. Lautenbach and R. Grauer}

\title{An asymptotic-preserving five-moment two-species plasma model coupled to an external magnetohydrodynamic solver}
\author{Magnus Deisenhofer\corresp{\email{magnus.deisenhofer@ruhr-uni-bochum.de}}, Aleksandr Mustonen, Simon Lautenbach and Rainer Grauer}
\affiliation{Institute for Theoretical Physics I, Ruhr University Bochum,
Bochum, Germany}

\begin{document}

\maketitle

\begin{abstract}
Accurately modeling collisionless space plasmas requires capturing small-scale kinetic effects while keeping global-scale simulations computationally tractable. Traditional multiscale approaches often rely on localized magnetohydrodynamics (MHD)-particle-in-cell (PIC) coupling or dynamic model hierarchies. In this work, we extend an established, adaptive multi-model hierarchy spanning from fully kinetic Vlasov descriptions to fluid models by introducing an asymptotic-preserving (AP) strategy that couples a two-species, five-moment fluid description with an ideal MHD solver. This coupling is the final critical step toward enabling efficient global simulations because the kinetic-scale physics in nonideal regions is entirely handled by finer models in the hierarchy. Kinetic descriptions natively solve Maxwell's equations and thus capture fast plasma waves, oscillations, and light waves, which are absent in the  MHD dynamics. To address this difference without sacrificing computational efficiency, our AP framework seamlessly projects these fast dynamics onto the slow MHD dynamics, ensuring rigorous consistency at the model interfaces. We detail the AP two-fluid formulation, the variable-coupling interface, and its integration into external frameworks. Finally, we demonstrate the validity and robustness of the fully coupled framework, from kinetics to ideal MHD, through magnetic reconnection simulations.
\end{abstract}

\keywords{
multiscale multiphysics coupling,
asymptotic-preserving methods,
collisionless space plasmas,
magnetic reconnection}

\section{Introduction}

In the study of collisionless space plasmas, accurate modeling is crucial for understanding plasma dynamics across different scales. While the Vlasov equation provides a comprehensive kinetic description of plasma behavior, its high computational cost makes it impractical for simulating large global regions. Consequently, simplified models like Magnetohydrodynamics (MHD) are often employed, where non-ideal plasma behavior is captured through a generalized Ohm's law. However, these models do not fully account for the multiscale nature of plasma dynamics, particularly the influence of small-scale kinetic effects and full electrodynamics on large-scale behavior.

In this work, we pursue a different approach by focusing on multiscale multiphysics simulations, which allow for the exploration of large-scale plasma behavior driven by small-scale kinetic processes. Substantial progress has already been made by coupling MHD with kinetic Particle-In-Cell (PIC) methods, as demonstrated in \cite{wang-chen-toth:2022a, shou-tenishev-chen-toth:2021,  markidis-olshevskyetal:2021}. An alternative adaptive approach, presented in \cite{rieke-trost-grauer:2015, lautenbach-grauer:2018, allmannrahn-lautenbach-deisenhofer-grauer:2024}, involves coupling a hierarchy of models with varying complexity based on criteria that dynamically determine the appropriate modeling. Building on this, we propose extending the adaptive approach with an asymptotic-preserving strategy to couple two-species, five-moment equations and Maxwell’s equations with an ideal MHD solver.

\begin{figure}
    \centering
    \includegraphics[width=\linewidth]{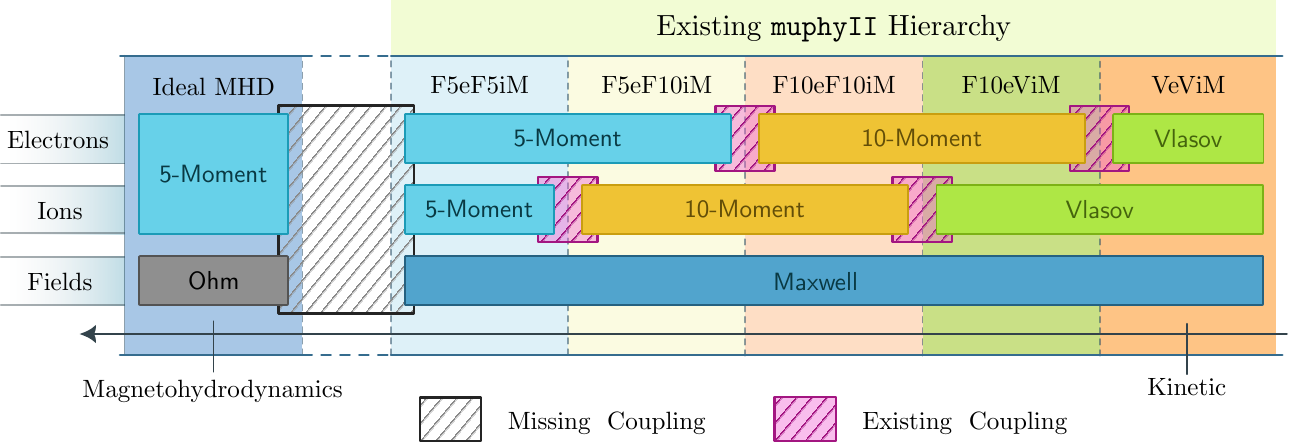}
    \caption{Current state of the scheme hierarchy implemented in \muphy in relation to ideal MHD, showing the models used for each plasma description and the required coupling.}
    \label{fig:hierarchy}
\end{figure}

For this approach, the coupling to an \textit{ideal} MHD solver is sufficient because all non-ideal regions are captured by a hierarchy of models that account for kinetic-scale physics.
The hierarchy described in \cite{allmannrahn-lautenbach-deisenhofer-grauer:2024} proceeds from fine- to coarse-grained modeling as follows: beginning with the most detailed fully kinetic Vlasov descriptions for both electrons and ions, continuing with a hybrid model comprising ten-moment electrons and Vlasov ions, followed by ten-moment treatments of both species, then a reduced description with five-moment electrons and ten-moment ions, and leading to five-moment models for both electrons and ions.
This structured approach depicted in Fig. \ref{fig:hierarchy} ensures that the appropriate level of detail is applied to different regions, allowing for efficient and accurate simulation of both ideal and non-ideal plasma behavior.

In this paper, we focus on the final crucial step towards enabling global simulations: coupling the two-species five-moment equations with the one-fluid MHD equations. In the two-fluid and kinetic models within the hierarchy, electromagnetic fields are governed by Maxwell's equations, while in the MHD system, the electric field is determined by an (ideal) Ohm's law (see also \cite{ho-datta-shumlak:2018}). Consequently, fast plasma waves and oscillations, and especially light waves, which are naturally present in kinetic models, are absent in the MHD description. As highlighted in \cite{burby:2017, miloshevich-burby:2021}, MHD dynamics takes place on a slow manifold, where the influence of fast dynamics is subdued. Asymptotic-preserving (AP) methods, as discussed in \cite{degond-deluzet-doyen:2017, crestetto-deluzet-doyen:2020}, offer a means to project the fast dynamics onto this slow manifold, thus ensuring consistency between the kinetic and MHD models while preserving computational efficiency.

The outline of this paper for coupling an asymptotic-preserving two-species five-moment model to ideal MHD is as follows:
Section \ref{sec:AP} details the asymptotic-preserving two-species five-moment model coupled with Maxwell's equations. We derive the numerical scheme that projects fast dynamics onto the slow MHD dynamics, and describe the temporal discretization and the strategy for solving the arising implicit system.
Section \ref{sec:MHD-two-fluid-variables} describes the coupling algorithm between the five-moment solver and the ideal MHD solver. We discuss the interface conditions and the mechanisms for variable exchange.
Section \ref{sec:MuPhy-external} addresses the coupling of the \muphy framework to an external MHD solver. This implementation serves as a prototype for coupling to more sophisticated external (AMR) MHD solvers.
Section \ref{sec:application} presents numerical applications demonstrating the capabilities of the proposed scheme. We focus on magnetic reconnection scenarios to validate the coupled model and highlight the efficiency gains of the multiscale approach. Furthermore, we showcase a fully coupled simulation spanning the entire hierarchy from kinetic to MHD scales, illustrating the seamless transition between all model levels.
Finally, Section \ref{sec:Outlook} concludes the paper with a summary of our findings and outlines perspectives for future extensions of the framework.

\section{Asymptotic-Preserving Two Fluid Description}
\label{sec:AP}

A necessity in smooth transition between different spatial regions arises when dealing with large simulation domains. 
For example, accurate reconnection simulations require a model capable of describing all of the physical phenomena occurring inside the thin current sheet and in the closest neighboring regions, such as a direct description based on the Vlasov equation. However, if we want to depict the physics in the outer regions of the simulations, we can use faster and more computationally efficient models. These models include 10-moment, 5-moment, and MHD descriptions, arranged in this order by a decrease in both the computational cost required and number of physical phenomena they can describe respectively. 
The 10-moment fluid models, which are obtained by taking moments of the Vlasov equation, employ a specific closure for the energy transport equation  \citep{sulem-passot:2015,
wang-hakim-etal:2015,
allmann-rahn-trost-grauer:2018,
hunana-tenerani-Zank-etal:2019,
ng-hakim-etal:2020,
allmann_rahn-lautenbach-grauer-sydora:2021}
and, depending on the closure, can retain some of the kinetic effects related to the shape of the distribution function, like Landau damping \citep{lautenbach-luebke-etal:2026}.
In the case of 5-moment fluid models, isotropic pressure is assumed, which neglects all effects related to deviations from a Maxwellian distribution function and simplifies the expression for the moment equations to the Euler equations for each species \textit{s}:
\begin{alignat}{2}
\frac{\partial n_s}{\partial t}&+\nabla \cdot\left(n_s \mathbf{u}_s\right)&&=0 \label{eq:0m} \\
m_s \frac{\partial\left(n_s \mathbf{u}_s\right)}{\partial t}&+ \nabla \cdot \left(p_s\mathbf{I} + m_s n_s \mathbf{u}_s \otimes \mathbf{u}_s\right)&&=n_s q_s\left(\mathbf{E}+\mathbf{u}_s \times \mathbf{B}\right)\label{eq:1m}\\
\frac{\partial \mathcal{E}_s}{\partial t}&+ \nabla \cdot\left(\mathbf{u}_s\left(\mathcal{E}_s + p_s\right)\right)&&=q_s n_s \mathbf{u}_s \cdot \mathbf{E} \label{eq:2m}
\end{alignat}
where $\mathbf I$ is the identity matrix and $\otimes$ denotes the tensor product. The species are characterised by their mass $m_s$, their charge $q_s$ and their distribution function $f_s$, from which the fluid moments and the related macroscopic quantities density $n_s$, fluid velocity $\mathbf u_s$, scalar pressure $p_s$, and scalar energy density $\mathcal E_s$ are derived:
\begin{align}
    n_s &= \int f_s \mathrm{~d} v^N \\
    n_s \mathbf{u}_s &= \int \mathbf{v} f_s \mathrm{~d} v^N \\
    p_s &= \frac{m_s}{N} \int (\mathbf{v}-\mathbf{u}_s)^2 f_s \mathrm{~d} v^N\\
    \mathcal{E}_s &=\frac{m_s}{2} \int \mathbf{v}^2 f_s \mathrm{~d} v^N=\frac{N}{2}p_s+\frac{1}{2}m_s n_s \mathbf{u}^2_s \label{eq:moments-E}
\end{align}
Here, $N$ denotes the dimension of the velocity space.
Due to the absence of collisional effects, the interaction between ions and electrons is carried out via collective electromagnetic fields. The evolution if the electric field $\mathbf E$ and magnetic field $\mathbf B$ is described by Maxwell's equations in a vacuum
\begin{subequations}
\label{eq:maxwell}
\begin{flalign}
    &\text{Ampère law:}& \nabla \times \mathbf{B} &= \mu_0 \mathbf{j} + \frac{1}{c^2}\partial_t \mathbf{E}&&&\label{eq:ampere}\\
    &\text{Faraday law:}& \nabla \times \mathbf{E} &= - \partial_t\mathbf{B}&&&\label{eq:faraday}\\[3pt]
    &\text{Magnetic Gauss law:}& \nabla \cdot \mathbf{B} &= 0&&&\label{eq:magnetic-gauss}\\
    &\text{Electric Gauss law:}& \nabla \cdot \mathbf{E} &= \frac{\rho}{\varepsilon_0}&&&\label{eq:electric-gauss}
\end{flalign}
\end{subequations}
coupled to the fluid equations via the charge density $\rho$ and current density $\mathbf j$:
\begin{align}
    \rho &= \sum_s q_s n_s\\
    \textbf{j}&= \sum_s \textbf{j}_s = \sum_s q_s n_s \textbf{u}_s \label{eq:curr} 
\end{align}
For the rest of this discussion, we will be focusing on a system of two plasma species: electrons with $q_e = -e$ and positive single-charged ions with $q_i = e$.
\subsection{From the Two-Fluid model to ideal MHD}\label{sec:background}
The previously introduced two-fluid-Maxwell system can be reduced further towards Magnetohydrodynamics by eliminating small spatial scales and fast time scales from the model.
The MHD equations can be derived by assuming that the electron mass $m_e$ and the vacuum permittivity $\varepsilon_0$ are negligible.
The former causes certain electron scales, like the electron inertial length and gyroradius, to be small relative to the scales of our system, with the resulting model only resolving ion scales. The latter assumption eliminates further time and spatial scales related to the coupling to the electromagnetic fields, particularly the Debye length, the plasma periods and the light crossing time.
In the following, we apply the asymptotic approximations $m_e \rightarrow 0\ $ and $\ \varepsilon_0 \rightarrow 0$ to the above two-fluid-Maxwell system.
Maxwell's equations are only affected by the latter:
\begin{align}
    \varepsilon_0 \nabla \cdot \mathbf{E} &= \rho&& \ \xrightarrow{\ \varepsilon_0 \rightarrow 0\ }\ &\rho &= 0 \\
    \nabla \times \mathbf{B} &= \mu_0\mathbf{j} + \varepsilon_0 \mu_0 \partial_t \mathbf{E}&& \ \xrightarrow{\ \varepsilon_0 \rightarrow 0\ }\ &\nabla \times \mathbf{B} &= \mu_0\mathbf{j}\label{eq:modified-ampere}
\end{align}
The first limit eliminates the divergence constraint on $\mathbf{E}$ and directly implies quasi-neutrality. The second equation loses the term $\partial_t \mathbf{E}$ which is required for the derivation of both plasma oscillations and electromagnetic waves. 
With the time derivative of $\mathbf{E}$ no longer determined by the Ampère law, we need some alternative way of expressing how $\mathbf{E}$ evolves.
By combining the first fluid moment (\ref{eq:1m}) of the particle species, we can derive a relation between $\mathbf{j}$ and $\mathbf{E}$, along with other quantities, that is independent of $\varepsilon_0$.
In analogy to the classic Ohm's law for electrical circuits, which similarly relates current and electric field, this relation is known as a generalized Ohm's law.
By multiplying (\ref{eq:1m}) with the species' charge, we get a relation between the single species currents $\mathbf{j}_s = q_s n_s \mathbf{u}_s$ and $\mathbf{E}$:
\begin{align}
    \partial_t \mathbf{j}_s + \frac{q_s}{m_s}\nabla p_s + q_s\nabla \cdot (n_s \mathbf{u}_s \otimes \mathbf{u}_s) = \frac{q_s^2}{m_s}n_s\mathbf{E} + \frac{q_s}{m_s} \mathbf{j}_s \times \mathbf{B}\label{eq:single_species_ohm}
\end{align}
To obtain the generalized Ohm's law for a combined system of electrons and ions, we can combine the equations for each species:
\begin{align}\label{eq:ohm-law-general}
    &\ e^2(n_e + \frac{m_e}{m_i}n_i)\mathbf{E} + e^2(n_e\mathbf{u}_e + \frac{m_e}{m_i}n_i\mathbf{u}_i) \times \mathbf{B} \nonumber\\ = &\ m_e\partial_t  (\mathbf{j}_i + \mathbf{j}_e) + e\nabla (\frac{m_e}{m_i}p_i - p_e) + em_e \nabla \cdot (n_i \mathbf{u}_i \otimes \mathbf{u}_i - n_e \mathbf{u}_e \otimes \mathbf{u}_e)
\end{align}
This is the most general form of the generalized Ohm's law that also applies to the regular Vlasov-Maxwell system, though it may additionally contain a resistive term originating from the collision operator which is omitted in our case.
Due to quasi-neutrality, we can replace the densities by a single common density $n$:
\begin{align}
    \rho = e(n_i - n_e) = 0\ \ \Rightarrow\ \ n_e = n_i = n
\end{align}
Due to our first asymptotic approximation $m_e \rightarrow 0$, the general Ohm's law loses the terms relating to the electron inertia:
\begin{alignat}{2}
    &&\ e^2n\mathbf{E} + e^2n\mathbf{u}_e \times \mathbf{B} &= -e\nabla p_e\\
    \Leftrightarrow &&\ e^2n\mathbf{E} + e^2(n\mathbf{u}_i - \mathbf{j} / e) \times \mathbf{B} &= -e\nabla p_e\\
    \Leftrightarrow &&\ \mathbf{E} + \mathbf{u}_i \times \mathbf{B} &= \frac{\mathbf{j} \times \mathbf{B}}{ne} -\frac{\nabla p_e}{ne}\label{eq:hall}
\end{alignat}
This form of the Ohm's law is the starting point of Hall-MHD. 
Assuming that the current component perpendicular to the magnetic field is small compared to that of the fluid velocity
\begin{align}
    |\mathbf{j} \times \mathbf{B}|/en = |(\mathbf{u}_i - \mathbf u_e)\times \mathbf{B}| \ll |\mathbf{u} \times \mathbf{B}|\label{eq:ohm-assumption}
\end{align} 
where $\mathbf u$ is the centre-of-mass fluid velocity defined in \eqref{equ:com_rhou}, 
and that the term due to the electron pressure is negligible
\begin{align}
    |\nabla p_e|/en \ll |\mathbf{E}|
\end{align}
the Ohm's law can be further approximated by the Ohm's law of ideal MHD:
\begin{align}
    \mathbf{E} + \mathbf{u}_i \times \mathbf{B} = 0 \label{eq:ideal-proto}
\end{align}

The multi-fluid description can be combined into a single-fluid description by adding up the moment equations (\ref{eq:0m}-\ref{eq:2m}).
In general, this combined system is not closed, as the equations for the combined fluid variables may still depend on the fluid variables of the individual original fluids.
In the special case of ideal MHD, this combined fluid description can be written down as a single set of 5-moment equations, which, together with (\ref{eq:modified-ampere}), form a conservation law.
To that end, we introduce the center of mass variables:
\begin{align}
    \rho &= m_en_e + m_in_ii\label{equ:com_rho}\\
    \rho\mathbf{u} &= m_en_e\mathbf{u}_e + m_in_i\mathbf{u}_i  \label{equ:com_rhou}\\
     p &= p_e + p_i \label{equ:com_p}\\
     \mathcal{E} &= \mathcal{E}_e + \mathcal{E}_i + \frac{1}{2\mu_0}|\mathbf{B}|^2\label{equ:com_eps}
\end{align}
In the limit $m_e \rightarrow 0$, the density and momentum variables reduce to the ion ones:
\begin{alignat}{2}
    \rho &= m_en_e + m_in_i &&\ \xrightarrow{m_e\ \rightarrow\ 0}\ m_in_i\\
    \rho\mathbf{u} &= m_en_e\mathbf{u}_e + m_in_i\mathbf{u}_i  &&\ \xrightarrow{m_e\ \rightarrow\ 0}\ m_in_i\mathbf{u}_i
\end{alignat}
As a consequence, the center of mass velocity reduces to the ion velocity
\begin{align}
    \mathbf{u} = \frac{m_e n_e \mathbf{u}_e + m_i n_i \mathbf{u_i}}{m_e n_e + m_i n_i}\ \xrightarrow{m_e\ \rightarrow\ 0}\ \mathbf{u}_i
\end{align}
so that the ideal Ohm's law is:
\begin{align}
    \mathbf{E} = - \mathbf{u} \times \mathbf{B}\label{eq:ideal-ohm}
\end{align}
While this choice of variables is common, it is not unique in the sense that other definitions with the same limit for $m_e \rightarrow 0$ will also produce the ideal MHD equations. The variable definitions before the limit present just one possible way of relating the MHD quantities to those of a finite mass ratio, two-fluid system to approximate.

By taking a mass-weighted sum of the five-moment equations, inserting the ideal Ohm's law into the source terms and taking the two asymptotic limits, the system can be rewritten in the form of the following conservation law, which is commonly implemented in MHD codes:
\begin{align}
    \partial_t
    \begin{pmatrix}
    \rho \\ \rho \mathbf{u} \\ \mathcal{E} \\ \mathbf{B}
    \end{pmatrix} + \nabla \cdot &
    \begin{pmatrix}
        \rho \mathbf{u} \\
        \rho \mathbf{u} \otimes \mathbf{u} + (p + \frac{1}{2\mu_0}|\mathbf{B}|^2)\mathbf{I} - \frac{1}{\mu_0}\mathbf{B} \otimes \mathbf{B} \\
        \mathbf{u} (\mathcal{E} + p + \frac{1}{2\mu_0}|\mathbf{B}|^2) - \frac{1}{\mu_0}\mathbf{B}(\mathbf{u}\cdot\mathbf{B})\\
        \mathbf{u} \otimes \mathbf{B} - \mathbf{B} \otimes \mathbf{u}
    \end{pmatrix} = 0\label{eq:mhd-conservative}\\
    \mathcal{E} &= \frac{N}{2} p + \frac{1}{2}\rho|\mathbf{u}|^2 + \frac{1}{2\mu_0}|\mathbf{B}|^2
\end{align}

\subsection{Asymptotic Preserving discretisation of Maxwell's equations}

All the different schemes currently implemented in \muphy \citep{allmannrahn-lautenbach-deisenhofer-grauer:2024}, ranging from Vlasov to the five-moment model, are coupled to Maxwell's equation.
Magnetohydrodynamics makes the approximation of an infinite speed of light, since taking the limit $\varepsilon_0 \rightarrow 0$, while keeping $\mu_0$ constant, implies $c\rightarrow \infty$, which is a significant difference in physics compared to the Maxwell system.

The goal of this section is to derive a method of solving for the electromagnetic fields that can be tuned to be consistent with either Maxwell's equations or a generalised Ohm's law using $\varepsilon_0$ as a parameter. 
One potential approach involves utilizing multiple blocks with varying values of $\varepsilon_0$ to distribute the transition from a non-zero to a zero permittivity over a larger spatial distance, thereby mitigating potential boundary artifacts arising from the discontinuity. However, this staggered strategy proves to be unnecessarily complex, as executing the transition in a single step yields stable results without introducing adverse boundary effects.

As discussed in the derivation of MHD, in this transition, the Maxwell-Euler system lacks an equation to determine the evolution of the electric field
\begin{align}
    \nabla \times \mathbf{B} = \mu_0\mathbf{j} + \varepsilon_0 \mu_0 \partial_t \mathbf{E} \ \xrightarrow{\ \varepsilon_0\ \rightarrow\ 0\ }\ \nabla \times \mathbf{B} = \mu_0\mathbf{j}
    \nonumber
\end{align}
and needs an additional equation in the form of an Ohm's law to compute $\mathbf{E}$.
For us to be able to freely choose $\varepsilon_0$ without having to switch to a different numerical method close to $\varepsilon_0 = 0$, we base our scheme off of a single equation that provides a means to compute $\mathbf{E}$ within the entire parameter range.

Any numerical scheme which is consistent with Maxwell's equations and has a stability condition dependent on the speed of light, like most explicit methods, will have asymptotically small maximum time steps close to $\varepsilon_0 = 0$.
The best candidates for schemes without such a constraint are implicit methods whose stencil covers the entirety of its domain. 
Another motivation for using an implicit solver is that under-resolving oscillations or waves using an implicit solver tends to lead to damping, which may help dampen electromagnetic waves before they can reach the boundary and be reflected.
A downside of such implicit schemes is that they can potentially be very expensive to solve, since they may require a large number of iterations to converge to a solution.
Thus, another desired property of our scheme is that the implicitly discretised system can be solved efficiently.
The properties we want our numerical scheme to have are the defining properties of so-called Asymptotic-Preserving schemes, as coined by \cite{Jin1999}.

Our derivation of an asymptotic preserving scheme largely follows the approach in \cite{degond-deluzet-doyen:2017}, who present a way of deriving a scheme with the properties we are looking for, which is implicit and first order accurate in time. 
Throughout their derivation, they use a normalised form of the Vlasov-Maxwell system and discuss the limiting behaviour of a parameter $\lambda$, which represents the ratio of the Debye length $\lambda_D$ to some reference system scale.
Due to the various constraints placed on the coefficients of this normalised system, the limit $\lambda \rightarrow 0$ is equivalent to $\varepsilon_0 \rightarrow 0$ applied to our unnormalised system.
The derivation only considers the quasi-neutral limit and not the mass limit, as the latter does not directly affect the electromagnetic fields. 
Their solver is designed to be integrated into a PIC solver, which represents the distribution functions using a collection of macro-particles.
For our use case, we want to derive a similar scheme suited for use with our existing finite volume solver for the fluid variables.
Since the original paper does not discuss how to numerically solve the discrete linear system resulting from the discretisation, we also present an efficient method for calculating a solution.

The general strategy is to first derive a reformulated version of Ampère's law, which provides an equation for $\mathbf{E}$ regardless of the value of $\varepsilon_0$, and then find a semi-discretisation for Maxwell's equations that is consistent with that form of Ampère's law.
The reformulated Ampère law is obtained by combining the two different expressions for the time derivative of the total current resulting from Maxwell's equations and from the generalised Ohm's law \eqref{eq:single_species_ohm}:
\begin{align}
    \partial_t \mathbf{j} &= - \frac{1}{\mu_0} \nabla \times \nabla \times \mathbf{E} - \varepsilon_0 \partial_t^2 \mathbf{E}\\
    \partial_t \mathbf{j} &= \sum_s \partial_t \mathbf{j}_s = \sum_s \left[ \frac{q_s^2}{m_s}n_s\mathbf{E} + \frac{q_s}{m_s}\mathbf{j}_s \times \mathbf{B} - \frac{q_s}{m_s}\nabla \cdot \mathcal{P}_s\right]
\end{align}
where $\mathcal{P}_{s}=p_{s} \mathbf{I}+m_{s} n_{s} \mathbf{u}_{s} \otimes \mathbf{u}_{s}$.
Setting both right-hand sides equal and moving all terms involving $\mathbf{E}$ to one side gives us our desired equation:
\begin{align}
    \varepsilon_0 \partial_t^2 \mathbf{E} + \sum_s \frac{q_s^2}{m_s}n_s\mathbf{E} + \frac{1}{\mu_0}\nabla \times \nabla \times \mathbf{E} = \sum_s \frac{q_s}{m_s}\left[ \nabla \cdot \mathcal{P}_s - \mathbf{j}_s \times \mathbf{B}\right]\label{eq:reformulated-ampere}
\end{align}

Instead of discretising the reformulated equation directly, \citeauthor{degond-deluzet-doyen:2017} instead choose to modify a discretisation of the standard Maxwell system that is consistent with it.
To ensure that there are no time step size restrictions, they use a modified version of the $\theta$-method for the time discretisation.
The $\theta$-method is an implicit Runge-Kutta method and a generalisation of the forward and backward Euler methods, which is A-stable for $\theta \in [\frac12,1]$.
For an ordinary differential equation of the form $\partial_t y(t) = f(y, t)$ the $\theta$-method computes an approximation for $y^{n+1} = y(t^{n+1} = t^n + \Delta t)$ from $y^n = y(t^n)$ as follows:
\begin{align}
    y^{n+1} &= y^n + \Delta t \left[(1 - \theta) f(y^n, t^n) + \theta f(y^{n+1}, t^{n+1})\right]
\end{align}
This formula can be applied directly to the Maxwell system \eqref{eq:maxwell}:

\begin{align}
    \varepsilon_0\mathbf{E}^{n+1} &= \varepsilon_0\mathbf{E}^n + \Delta t\Big[\frac{1}{\mu_0} \nabla \times \mathbf{B}^{n + \theta} - \sum_s \mathbf{j}_s^{n+\theta}\Big]\label{eq:ap-E}\\
    \mathbf{B}^{n+1}&= \mathbf{B}^n - \Delta t \Big[\nabla \times \mathbf{E}^{n+\theta}\Big]\label{eq:ap-B}
\end{align}
Here we use the superscript $n + \theta$ to refer to linear combinations of the following form:
\begin{align}
    \mathbf{E}^{n+\theta} = (1-\theta)\mathbf{E}^{n} + \theta \mathbf{E}^{n+1}
\end{align}
To make this method consistent in the quasi-neutral limit, we need a term involving $\mathbf{E}^{n+1}$ that does not vanish for $\varepsilon_0 \rightarrow 0$.
The proposed solution to this is to replace $\mathbf{j}^{n+\theta}$ by an approximation of $\mathbf{j}^{n+1}$, using the generalised Ohm's law to write it in the form:
\begin{align}
    \mathbf{j}^{n+1}_s = \mathbf{j}_s^{n+1,*} + \Delta t C_s \mathbf{E}^{n+1}\label{eq:j-approximation}
\end{align}
Comparing this with (\ref{eq:single_species_ohm}), we choose $C_s = n_s^{n+1} q_s^2 / m_s $ and 
\begin{align}
    \mathbf{j}_s^{n+1,*} = F\Big(\mathbf{j}_s^n, \frac{q_s}{m_s}\big[\mathbf{j}_s \times \mathbf{B} - \nabla \cdot \mathcal{P}_s\big], \Delta t\Big) 
\end{align}
where $F(y^n, f(y, t), \Delta t)$ is some numerical integration or approximation technique. 
\citeauthor{degond-deluzet-doyen:2017}'s modified $\theta$-method is constructed by inserting this Ansatz into (\ref{eq:ap-E}):
\begin{align}
    \left(\varepsilon_0 + \Delta t^2\sum_s \frac{q_s^2}{m_s}n_s^{n+1}\right)\mathbf{E}^{n+1} &= \varepsilon_0\mathbf{E}^n + \Delta t\Big[\frac{1}{\mu_0} \nabla \times \mathbf{B}^{n + \theta} - \sum_s \mathbf{j}_s^{n+1,*}\Big]\label{eq:ap-modified-E}\\
    \mathbf{B}^{n+1}&= \mathbf{B}^n - \Delta t \Big[\nabla \times \mathbf{E}^{n+\theta}\Big]\label{eq:ap-modified-B}
\end{align}
Importantly, they show that this approach leads to a consistent discretisation of (\ref{eq:reformulated-ampere}) for both $\varepsilon_0 \neq 0$ and $\varepsilon_0 = 0$, given a suitable spatial discretisation, as long as the approximation for $\mathbf{j}^{n+1,*}$ is consistent.

\citeauthor{degond-deluzet-doyen:2017} present two different ways of approximating this term for PIC codes.
The second method described in the paper, the \enquote{AP-Particle scheme}, approximates $\mathbf{j}_s^{n+1,*}$ by running a particle update without any contribution from $\mathbf{E}$ and then calculating the resulting current. The values of $\mathbf{E}^{n+1}$ and $\mathbf{B}^{n+1}$, calculated by the AP solver, are then used to re-run the particle update.

For our fluid solver, we can implement a similar approximation by employing first order splitting between the source terms involving $\mathbf{E}$ and the rest of the five-moment equations. The first half of the split system is:
\begin{alignat}{2}
    \partial_t n_s &+ \nabla \cdot (n_s\mathbf{u}_s) &&= 0\\
    \partial_t (n_s\mathbf{u}_s) &+ \nabla \cdot (\frac{p_s}{m_s}\mathbf{I} + n_s\mathbf{u}_s \otimes \mathbf{u}_s) &&= \frac{q_sn_s}{m_s}(\mathbf{u}_s \times \mathbf{B})\\
    \partial_t \mathcal{E}_s &+ \nabla \cdot \mathbf{u}_s(\mathcal{E}_s + p_s) &&= 0
\end{alignat}
The second half just consists of the missing source terms:
\begin{align}
    \partial_t (n_s\mathbf{u}_s) &= \frac{q_s}{m_s}n_s\mathbf{E}\\
    \partial_t \mathcal{E}_s &= q_sn_s(\mathbf{u}_s \cdot \mathbf{E})
\end{align}
The first part can be solved with an existing finite volume solver for the five-moment equations by setting $\mathbf{E} = 0$, which has the advantage of retaining the good conservation properties of the solver for most of the update.
This first update computes the final density $n_s^{n+1}$, and the intermediate second moment $(n_s\mathbf{u}_s)^{n+1,*}$ and energy $\mathcal{E}_s^{n+1,*}$. The intermediate current $\mathbf{j}_s^{n+1,*}$ is then directly available by multiplying by the charge:
\begin{align}
    \mathbf{j}_s^{n+1,*} = q_s (n_s\mathbf{u}_s)^{n+1,*}
\end{align}
Stepping the source term using a simple Euler step ensures that the contribution of $\mathbf{E}^{n+1}$ to the current is consistent with the one that was used to calculate it in (\ref{eq:j-approximation}):
\begin{align}
    (n_s\mathbf{u}_s)^{n+1} &= (n_s\mathbf{u}_s)^{n+1,*} + \Delta t \frac{q_s}{m_s}n^{n+1}_s\mathbf{E}^{n+1}\\
    \mathcal{E}_s^{n+1} &= \mathcal{E}_s^{n+1,*} + \Delta t q_sn_s^{n+1}(\mathbf{u}_s^{n+1,*} \cdot \mathbf{E}^{n+1})
\end{align}

\citeauthor{degond-deluzet-doyen:2017} directly solve (\ref{eq:ap-modified-E} - \ref{eq:ap-modified-B}), but the system can be put into a more convenient form where $\mathbf{B}^{n+1}$ can be calculated explicitly after solving an implicit equation for $\mathbf{E}^{n+1}$.
First, we rewrite the system of equations in terms of source terms and a two by two matrix of linear operators 
\begin{align}
    \begin{pmatrix} \mathbf{D} & \mathbf{P} \\ \mathbf{Q} & 1\end{pmatrix} \begin{pmatrix}  \mathbf{E}^{n+1}\\\mathbf{B}^{n+1}\end{pmatrix} = \begin{pmatrix}
    \mathbf{u}\\
    \mathbf{v}
    \end{pmatrix}\label{eq:short_schur}
\end{align}
which is a handier shorthand for the full expression:
\begin{align*}
    \begin{pmatrix} 1 + \frac{\Delta t^2}{\varepsilon_0}\sum_s \frac{q_s^2 n_s}{m_s} & -c^2 \Delta t \theta \nabla \times \\ \Delta t \theta \nabla \times & 1\end{pmatrix} \begin{pmatrix}  \mathbf{E}^{n+1}\\\mathbf{B}^{n+1}\end{pmatrix} = \begin{pmatrix}
    \mathbf{E}^n + c^2\Delta t  (1 - \theta) \nabla \times \mathbf{B}^n-\frac{\Delta t}{\varepsilon_0} \sum_s \mathbf{j}_s^{n+1,*}\\
    \mathbf{B}^n-\Delta t (1 - \theta)\nabla \times \mathbf{E}^n
    \end{pmatrix}\label{eq:long_schur}
\end{align*}
The form of this matrix enables us to use a special case of a Schur Complement to reduce this system to a single implicit equation for $\mathbf{E}^{n+1}$.
Note that, because the bottom right entry of (\ref{eq:short_schur}) is a scalar, $\mathbf{B}^{n+1}$ can be written explicitly in terms of $\mathbf{E}^{n+1}$ and $\mathbf{v}$:
\begin{align}
    \mathbf{B}^{n+1} = \mathbf{v} - \mathbf{Q}\mathbf{E}^{n+1}
\end{align}
Inserting this into the first row of (\ref{eq:short_schur}) eliminates $\mathbf{B}^{n+1}$:
\begin{align}
    \mathbf{D}\mathbf{E}^{n+1} + \mathbf{P}(\mathbf{v} - \mathbf{Q}\mathbf{E}^{n+1}) = \mathbf{u}
    \ \ \Leftrightarrow\ \  (\mathbf{D} - \mathbf{P}\mathbf{Q})\mathbf{E}^{n+1} = \mathbf{u} - \mathbf{P}\mathbf{v}
\end{align}
Solving the linear system thus reduces to inverting the following operator:
\begin{align}
    \mathbf{T} := \mathbf{D} - \mathbf{P}\mathbf{Q} = \left(1 + \frac{\Delta t^2}{\varepsilon_0}\sum_s \frac{q_s^2 n^{n+1}_s}{m_s}\right) + \Delta t^2 c^2 \theta^2 \nabla \times \nabla \times\label{eq:ap-operator}
\end{align}
Once it has been fully discretised, this linear operator allows for efficient numerical inversion using iterative methods, such as the Conjugate Gradient method.  

All numerical tests in \cite{degond-deluzet-doyen:2017} use $\theta = 1$ in which case $\mathbf{j}_s^{n+\theta} = \mathbf{j}_s^{n+1}$ and the method is equivalent to the unmodified $\theta$-method using the approximation (\ref{eq:j-approximation}). 
For lower values of $\theta$, a qualitative difference between the two methods can be illustrated by looking at the extreme case of $\theta = 0$ where the inversion of (\ref{eq:ap-operator}) can be done explicitly.
While equations (\ref{eq:ap-E} - \ref{eq:ap-B}) reduce to a standard explicit Euler discretisation for (\ref{eq:ampere}) and (\ref{eq:faraday}), the AP scheme (\ref{eq:ap-modified-E} - \ref{eq:ap-modified-B}) retains a damping factor on the electric field because it considers the response of the current irrespective of $\theta$:
\begin{align}
    \mathbf{E}^{n+1} = \underbrace{\left(1 + \frac{\Delta t^2}{\varepsilon_0} \sum_s\frac{q_s^2 n_s^{n+1}}{m_s}\right)^{-1}}_{\text{Damping factor}} \underbrace{\left[\mathbf{E}^n + \Delta t\ c^2\nabla \times \mathbf{B}^{n} - \frac{\Delta t}{\varepsilon_0} \sum_s \mathbf{j}_s^{n+1, *}\right]}_{\text{Explicit Euler}}\label{eq:explicit_ap}
\end{align}
This also suggests that the first term of (\ref{eq:ap-operator}) is related to the damping of the electric field and that larger time steps and densities are expected to lead to stronger damping.

\section{Coupling MHD and two-fluid variables}
\label{sec:MHD-two-fluid-variables}

Coupling two different models requires some way of converting the physical quantities on both sides of the model boundary into variables of the other model.
For two-fluid to MHD coupling, we can use the center of mass variables (\ref{equ:com_rho} - \ref{equ:com_eps}) from section \ref{sec:background}:
\begin{subequations}
\label{eqs:mhd-variables}
\begin{align}
    \mathbf{B}_\text{MHD} &= \textbf{B}\\
    \rho_\text{MHD} &= n_e m_e + n_i m_i\\
    \rho_\text{MHD} \textbf{u}_\text{MHD} &= n_e m_e \mathbf{u}_e + n_i m_i \mathbf{u}_i\\
    \mathcal{E}_\text{MHD} &= \mathcal{E}_e + \mathcal{E}_i + \frac{1}{2}\mathbf{B}^2\label{eq:mhd-variables-E}
\end{align}
\end{subequations}
The above choice of MHD variables ensures that the total particle energy, mass density, and momentum are conserved when converting between the two sets of variables using the original $m_e$. The total energy is not conserved, since $\mathbf{E}$ is not present in the conversion at all.

In fact, as MHD omits several processes present in the full system, such as light waves and plasma oscillations, and instead describes the slow, large-scale dynamics of the plasma, the conversion should not attempt to conserve the total energy. These omitted processes are primarily associated with electric-field fluctuations and their contribution to the total energy, which are absent in MHD. Therefore, we do not consider the conservation of total energy to be a desirable property.

Relating the variables of the two continuous systems does not yet fully determine how the discretised systems are coupled.
\cite{ho-datta-shumlak:2018} propose two ways of coupling MHD variables to two-fluid ones, in the context of a Discontinuous Galerkin method. 
The first is a conservative coupling, that uses variable conversion with finite $m_e$ to calculate fluxes based on the two-fluid equations and summing them to construct the MHD fluxes.
However, this construction also means that non-ideal terms of the two-fluid equations contribute to the fluxes on the MHD side of the boundary.
In their tests, this coupling develops oscillations on the MHD side of the boundary when the MHD assumptions are violated on the two-fluid side.
The second calculates separate fluxes for both sides, with an asymptotic approximation to the variable conversion with $m_e \ll m_i$ on the MHD side, and, importantly, does not include the two-fluid electric field in the flux calculations of the MHD side.
This coupling leads to much smaller oscillations in their test.

We choose an approach similar to the latter, by calculating separate fluxes using the regular flux reconstruction of each solver applied to the converted boundary data, since we expect the inclusion of a non-ideal $\mathbf{E}$ to be the deciding factor in these boundary effects.
It also does not require modifying the time integration of either solver and allows for coupling schemes that do not use numerical fluxes, such as Finite Element or Finite Difference methods.
In contrast, we do not approximate the variable conversion, to preserve the previously mentioned conservation properties.
As mentioned by \citeauthor{ho-datta-shumlak:2018}, calculating separate fluxes means that the sum of each model's conservation variables is not necessarily conserved exactly, but as long as the assumptions of ideal MHD apply, they should be conserved approximately.

The two-fluid densities and velocities can in turn be obtained from the MHD ones from their definitions \eqref{eqs:mhd-variables} under the assumption of quasi-neutrality:
\begin{align}
    n_e = n_i &= \rho_\text{MHD} / (m_e + m_i)\\
    n_e\mathbf{u}_e &= \frac{\rho_\text{MHD} \textbf{u}_\text{MHD} - \mathbf{j}_\text{MHD} m_i / e }{m_e + m_i}\\
    n_i\mathbf{u}_i &= \frac{\rho_\text{MHD} \textbf{u}_\text{MHD} + \mathbf{j}_\text{MHD} m_e / e }{m_e + m_i}
\end{align}
The electric field is determined by the ideal Ohm's law:
\begin{align}
    \mathbf{E} = -\mathbf{u}_\text{MHD}\times \mathbf{B}
\end{align}
Calculating the two-fluid energies $\mathcal{E}_s$ from the MHD quantities requires some additional assumption about the two-fluid variables; to turn one energy variable into two.
In MHD derivations relying on high collisionality to justify using the Euler equations, the natural choice is assuming that temperatures of both species are equal to a common temperature $T$, since the high collisionality would also lead to the equalising of temperatures.
This assumption is used in \cite{ho-datta-shumlak:2018}.
More generally, the equal temperature assumption can be considered as a special case of a constant temperature ratio $\alpha_T = T_e / T_i$, which may be valid in more general cases.
In the case of a Maxwellian distribution function, the pressure is directly related to the temperature and density via the ideal gas law $p_s = n_s k_\text{B} T_s$, where $k_\text{B}$ is the Boltzmann constant. Plugging this into \eqref{eq:moments-E} lets us rewrite the energies in terms of temperatures and then relate these temperatures to the MHD energy using $\alpha_T$:
\begin{align}
    &&\mathcal{E}_s &= \frac{1}{2}m_s n \mathbf{u}_s^2 + \frac{N}{2}n k_B T_s \label{eq:def-Es}\\
    \Rightarrow&& \mathcal{E}_\text{MHD} &= \frac{1}{2} m_en\mathbf{u}_e^2 + \frac{1}{2} m_in\mathbf{u}_i^2 + \frac{N}{2}k_BnT_i(1 + \alpha_T) + \frac{1}{2}\mathbf{B}^2\\
    \Rightarrow&& k_BT_i &= \frac{1}{(1 + \alpha_T)}\underbrace{\frac{2}{Nn }\left[\mathcal{E}_\text{MHD} - \frac{1}{2} m_en\mathbf{u}_e^2 - \frac{1}{2} m_in\mathbf{u}_i^2 - \frac{1}{2}\mathbf{B}^2\right]}_{:=\ k_B T}
\end{align}
Here we introduce the temperature $T$ to express the total thermal energy of the MHD fluid obtained from subtracting the kinetic and magnetic energy terms from \eqref{eq:mhd-variables-E}.
This leads to the following energy coupling:
\begin{align}
    \mathcal{E}_e &= \frac{1}{2}m_e n \mathbf{u}_e^2 + \frac{\alpha_T}{1 + \alpha_T} \frac{N}{2}n k_B T\\ \mathcal{E}_i &= \frac{1}{2}m_i n \mathbf{u}_i^2 + \frac{1}{1 + \alpha_T} \frac{N}{2}n k_B T
\end{align}
A similar and related approach is taken in \cite{DALDORFF2014236}, where the total pressure $p_\text{MHD} = p_e + p_i$ is split back based on some constant factor $\alpha_p$ according to $p_e = \alpha_p\ p_\text{MHD}$ and $p_i = (1 - \alpha_p)\ p_\text{MHD}$. When setting $\alpha_T = \alpha_p / (1 - \alpha_p)$, this assumption is equivalent to the temperature ratio.
In realistic settings, especially within a large system, this ratio is not necessarily constant throughout the domain.

These conversions are consistent in the sense that converting variables from MHD to two-fluid and back results in the same values. The reverse is not true since there is necessarily a loss of information going from two fluids to one. 
If the assumptions of MHD and those used for the conversion are not true around the model boundary, this may lead to a discontinuity.

\section{Coupling to external frameworks}
\label{sec:MuPhy-external}

\definecolor{RUB-green}{HTML}{8dae10}
\definecolor{RUB-blue}{HTML}{17365c}
\definecolor{RUB-gray}{HTML}{e7e7e7}
\definecolor{RUB-red}{HTML}{e6332a}

\definecolor{RUB-darkgreen}{HTML}{819f11}

While coupling to an actual AMR code is outside the scope of this paper, we develop an algorithm to couple to a generic external solver and demonstrate it using a simple test program with static coupling.
The communication to and from the \muphy framework \citep{allmannrahn-lautenbach-deisenhofer-grauer:2024} is designed to be easily adaptable to any code that can provide boundary values for similar rectangular regions of its domain, and whose data can then be periodically replaced by results computed in \muphy. These could be sections of uniform or block-structured grids, but also rectangular cells with many degrees of freedom per cell, such as the ones found in Finite Element or Discontinuous Galerkin methods.

\begin{figure}
    \centering
    \includegraphics[width=0.9\textwidth]{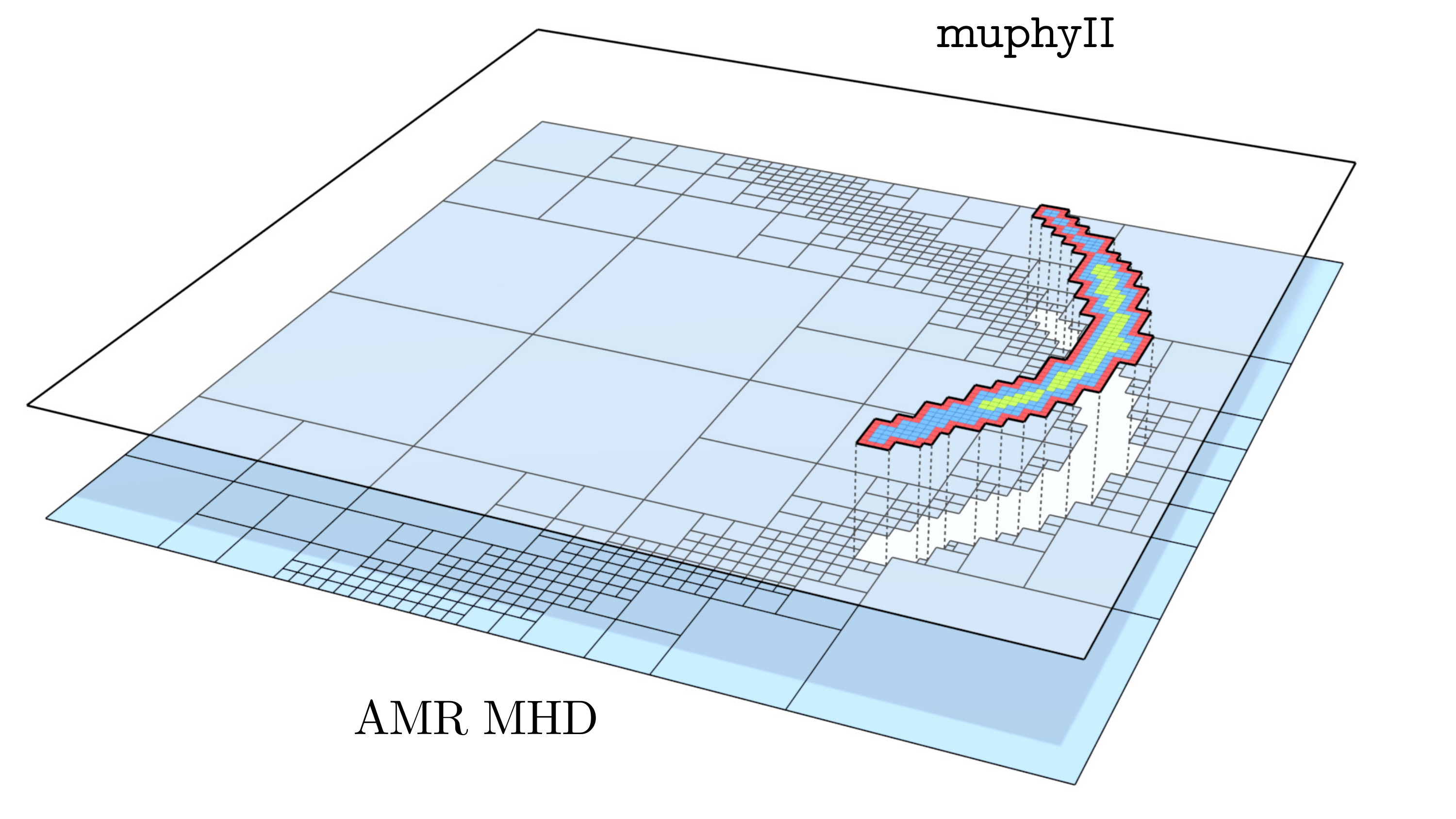}
    \caption{Sketch of our envisioned \muphy-AMR coupling. Both simulations span the whole domain, however only a part of the highest refinement level is being simulated by \muphy while the rest of the domain is exclusively simulated and stored by the external MHD code. A coupling layer (coloured red) at the edge of the \muphy region bridges the different electromagnetic models and time stepping approaches.}
    \label{fig:idea}
\end{figure}

The general premise is to select a sub-region of another program, ideally from the highest refinement level in the case of AMR, that has its calculations taken over by \muphy using a spatially overlapping set of blocks, such that the \muphy cells either map exactly to corresponding external cells, or there is some way to convert between the degrees of freedom for each physical quantity.
The \muphy grid covers the same domain as the external MHD solver, however inactive blocks do not perform calculations and do not store simulation data.
For a sketch of this idea, see Fig. \ref{fig:idea}.
For blocks on the boundary of the \muphy region, the corresponding data in the external program is overwritten with values converted from the two-fluid quantities after every time step. This avoids directly modifying the numerics of the external code.
Meanwhile, the ghost cells (or halo cells) of the sides of these boundary blocks facing the external region are filled by sampling the external data and converting it into two-fluid quantities.
To enforce that the boundary of the \muphy region of the domain is covered by \texttt{AP} scheme blocks, we introduce the \texttt{Empty} scheme, which represents blocks that are inactive on the \muphy side. \texttt{AP} sits between \texttt{Empty} and \texttt{F5eF5iM} in the hierarchy, automatically enforcing the boundary layer of \text{AP} blocks. 

Some specifics of the coupling process, such as the selection of blocks and the boundary value extraction, will depend on the choice of the external solver, however we can still discuss the general coupling strategy and the coupling internal to \muphy. 

In the previous discussion of the \texttt{AP} scheme we did not consider boundary conditions.
The operator that the AP solver numerically inverts operates on $\mathbf{E}^{n+1}$, so the appropriate Dirichlet boundary conditions are that of $\mathbf{E}^{n+1}$ from the neighbouring schemes.
This implies that the electromagnetic fields in the other schemes need to be computed first and then sent to the \texttt{AP} scheme.
In all existing schemes in \muphy (see Fig. \ref{fig:hierarchy}), Maxwell's equations are solved using a sub-cycled Finite-difference time-domain (FDTD) method where values are stored on a staggered grid and the time steps for the electric and magnetic fields are staggered in time. The electromagnetic solver is additionally offset in time relative to the fluid solver.
Since the electromagnetic fields of the \texttt{F5eF5iM} scheme are offset in time by half a time step with respect to the fluid variables, the FDTD solver will have to advance by only half a time step, before sending field data to the \texttt{AP} scheme and then waiting for that to return its results.

\begin{figure}
    \centering
    \includegraphics[width=\textwidth]{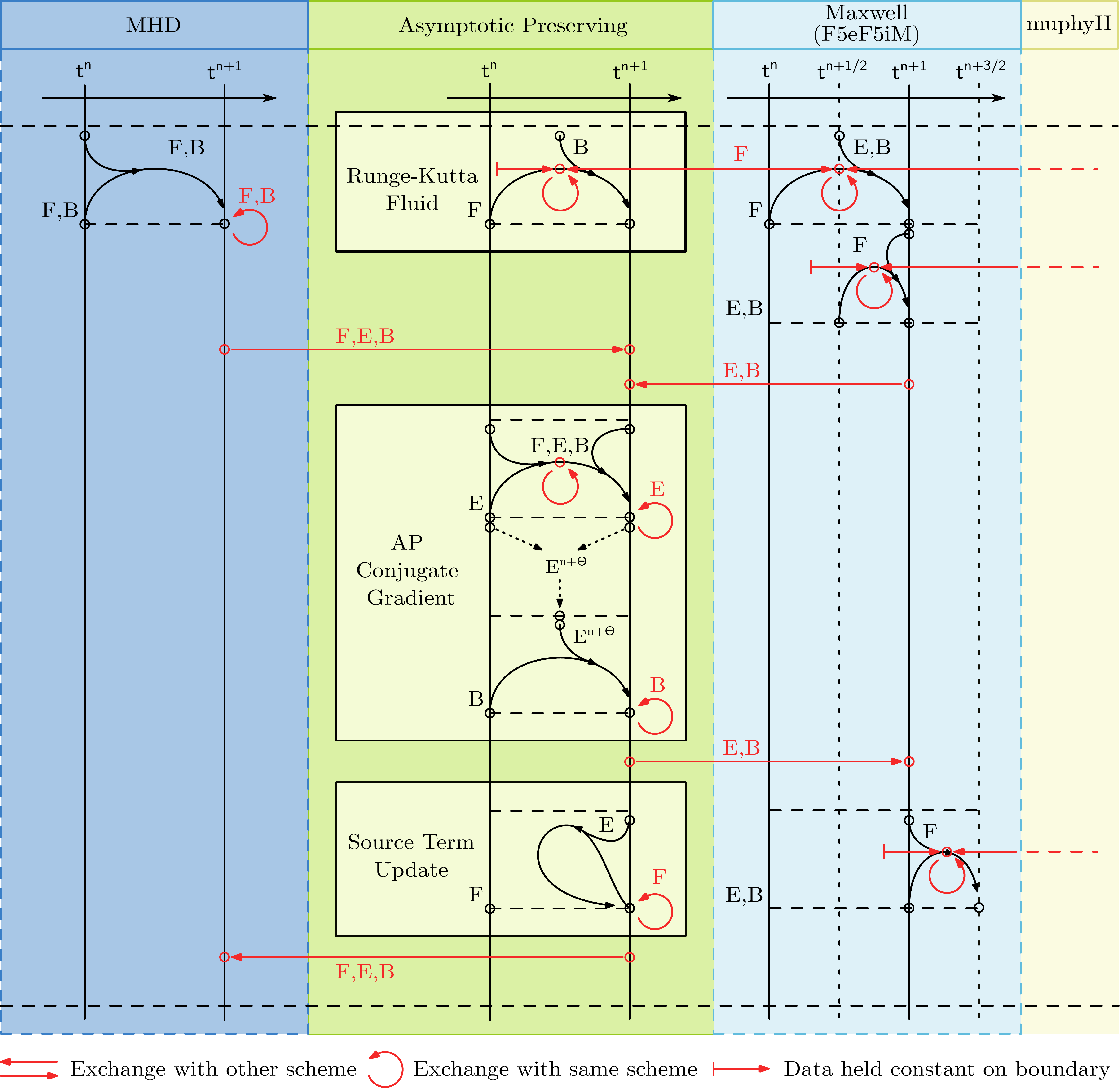}
    \caption{Logical steps in one time step of the AP scheme and neighbouring schemes, along with boundary data exchanges between them. The order of operations is from top to bottom. $\mathbf{F}$ denotes the set of fluid variables. Red arrows indicate boundary data exchange.}
    \label{fig:exchange_diagram}
\end{figure}

Some high level design decisions we are additionally following are to modify non-\muphy code as little as possible and to keep boundary values constant during steps that have sub-steps, such as the Runge-Kutta and sub-cycled FDTD  methods, on boundaries that face schemes that do not take the same sub-steps.
Fig. \ref{fig:exchange_diagram} illustrates the time stepping and data exchanges we propose to couple \texttt{AP}, \texttt{F5eF5iM} and the \texttt{External} solver, with a constant time step for clarity.

Not depicted in the diagram are time step calculations and the conversions that may be necessary before or after boundary data has been sent or received. 
In addition to conversion, it may be necessary to interpolate the converted quantities, as the external solver may use a different grid.
It is up to the specific implementation to perform conversion and interpolation either on the \muphy or the external side, whichever is more convenient or efficient.

We model the external solver as taking a step as described by Algorithm \ref{algo:mhd}.

\begin{algorithm}[H]
    \label{algo:mhd}
    \linespread{1.35}\selectfont
    \caption{External MHD Solver Step}
    
    \renewcommand{\SetProgSty}[1]{\renewcommand{\ProgSty}[1]{\textnormal{\csname#1\endcsname{##1}}\unskip}}
    \SetArgSty{text}
    \SetProgSty{textbf}
    
    \DontPrintSemicolon
    %\KwIn{List of clauses}
    %\KwOut{The clauses are satisfiable}
    \SetKwProg{Fn}{}{:}{}
    
    \Fn{External::step} {
        $(\rho^{n+1}, (\rho\mathbf{u})^{n+1}, \mathcal{E}^{n+1}, \mathbf{B}^{n+1}) \gets \textbf{Integrator}(\rho^{n}, (\rho\mathbf{u})^n, \mathcal{E}^{n}, \mathbf{B}^{n}, \Delta t)$\;
        $\Delta t \gets \textbf{Maxdt}(\rho^{n+1}, (\rho\mathbf{u})^{n+1}, \mathcal{E}^{n+1}, \mathbf{B}^{n+1})$\;
        find minimum $\Delta t$ among all blocks\;
    }
    
\end{algorithm}
Under the assumption that the optimal time step for the external solver is negotiated with \muphy, the changes needed are presented in Algorithm \ref{algo:modified_mhd}.

\begin{algorithm}[H]
    \label{algo:modified_mhd}
    \linespread{1.35}\selectfont
    \caption{Coupled External MHD Solver Step}
    
    \renewcommand{\SetProgSty}[1]{\renewcommand{\ProgSty}[1]{\textnormal{\csname#1\endcsname{##1}}\unskip}}
    \SetArgSty{text}
    \SetProgSty{textbf}
    
    \DontPrintSemicolon
    %\KwIn{List of clauses}
    %\KwOut{The clauses are satisfiable}
    \SetKwProg{Fn}{}{:}{}
    
    \Fn{External::step} {
        $(\rho^{n+1}, (\rho\mathbf{u})^{n+1}, \mathcal{E}^{n+1}, \mathbf{B}^{n+1}) \gets \textbf{Integrator}(\rho^{n}, (\rho\mathbf{u})^n, \mathcal{E}^{n}, \mathbf{B}^{n}, \Delta t)$\;
        $\Delta t \gets \textbf{Maxdt}(\rho^{n+1}, (\rho\mathbf{u})^{n+1}, \mathcal{E}^{n+1}, \mathbf{B}^{n+1})$\;
        \begingroup
        \color{RUB-darkgreen}
        find minimum $\Delta t$ among all blocks\;
        \uIf {this block is being simulated by \texttt{AP} scheme} 
        {
            send $\Delta t$ to \texttt{AP} scheme\;
            receive $\Delta t$ from \texttt{AP} scheme\;
        }
        \endgroup
        find minimum $\Delta t$ among all blocks\;
        \begingroup
        \color{RUB-darkgreen}
        \uIf {this block is being simulated by \texttt{AP} scheme} 
        {
            send $(\rho^{n+1}, \rho\mathbf{u}^{n+1}, \mathcal{E}^{n+1}, \mathbf{B}^{n+1})$\;
        
            receive $(\rho^{n+1}, \rho\mathbf{u}^{n+1}, \mathcal{E}^{n+1}, \mathbf{B}^{n+1})$ and place into ghost cells\;
        }
        \endgroup
    }
    
\end{algorithm}

Looking forward towards adaptive simulations, we would need additional functions to implement switching blocks between being simulated by \muphy and the external code.
These include functionality to determine which code should simulate which block, and to receive, send and convert block data.

\subsection{MPI Communication}\label{sec:mpi}

The data transfer between blocks in \muphy and subdivided domains in many other plasma codes is implemented using the Message Passing Interface (MPI), a standard for operations allowing communication between different processes (see \cite{mpi41}). 
This specification is implemented by multiple different libraries which allow applications to parallelise across machines and large compute clusters without having to implement the underlying network communication.

MPI works using the concept of communicators, a way of grouping processes and assigning them an identification number, known as a rank, to refer to by when sending data between them. 
By default, all initially launched processes are in a common \texttt{\detokenize{MPI_COMM_WORLD}} communicator.
Further communicators can be built by splitting it into subsets. 
Importantly, each process can be in multiple communicators simultaneously to use multiple sets of ranks to refer to other processes.

Since we want to couple two applications using MPI, an important concept is that of an intercommunicator.
Intercommunicators are a way for two disjoint communicators to refer to the processes in each other.
This has the advantage that each group of processes can work independently of the other within its own regular communicator and use the intercommunicator to address the other group by a separate set of ranks.
In our case, this enables us to add on remote communication without having to modify the existing communication logic of the individual codes.
We first split the processes into one communicator for each application, after which we construct an intercommunicator between them.
Most MPI implementations support Multiple Program Multiple Data (MPMD) launching, where multiple programs are started with a common \texttt{\detokenize{MPI_COMM_WORLD}} communicator.
For the \texttt{Open MPI} implementation (\cite{openmpi}), such a call would take the following form:
\begin{align}
    \texttt{\detokenize{mpirun [ options1 ] <program1> : [ options2 ] <program2> }}\nonumber
\end{align}

To determine which ranks in the intercommunicator to exchange data with, each application needs to implement a way of mapping MPI ranks of the other application to block coordinates.
For the simulations of this paper with static domain assignment, this translation can be worked out at compile time, however adaptive simulations will require communicating this information between the two codes.

\subsection{Test Program}

To test the validity of our approach and run coupled simulations, we implement a simple test program that takes the place of the external code.
The program implements the ideal MHD equations (\ref{eq:mhd-conservative}) using a block structured finite volume solver, written in python and parallelised using the \texttt{mpi4py} library (\cite{Dalcin2021}).
The numerical fluxes are evaluated using CWENO reconstruction (\cite{Kurganov2000}) and the time integration uses a third order Runge-Kutta method (\cite{SHU1988439}).
The overall time stepping is the same as the model of our external solver in Algorithm \ref{algo:mhd} and the additional communication functions are added as in Algorithm \ref{algo:modified_mhd}.
The time step size is taken as the minimum of that of the two codes and synchronised between them.
The blocks are of the same size and at the same locations as those of the \muphy grid; consequently, the mapping between block locations and ranks is the same as that of \muphy.
For boundary blocks, those simulated using the \texttt{AP} scheme, we send and receive the data of the entire block, converting and interpolating it into the correct ghost cells in \muphy, in particular for the staggered grid of the Maxwell solver.
The assignment of schemes to blocks, including which ones are active, is static in our test.

\section{Applications and Results}\label{sec:application}

This section presents numerical simulations of magnetic reconnection to validate the proposed coupling framework. In Sec.~\ref{sec:MHD-AP-explicit}, we examine the asymptotic-preserving (AP) coupling between the ideal MHD model and the AP solver. We also included the coupling from the AP scheme to the explicit five-moment scheme to demonstrate that no artificial effects arise on this side of the coupling. This simulation employs a reduced ion-to-electron mass ratio of $m_i/m_e = 25$. This choice is deliberate: a smaller mass ratio expands the spatial scale of kinetic-layer structures, thereby amplifying the numerical and physical challenges associated with transitions between model regimes. Crucially, when $m_{e} \nrightarrow 0$, the formal asymptotic limit from the two-fluid description to the MHD closure is no longer strictly satisfied. Consequently, the reduced-mass-ratio configuration provides a more stringent benchmark for the robustness and accuracy of the model coupling than simulations performed with physical mass ratios.

In Section \ref{sec:MHDto10moment}, we combine the AP coupling strategy presented here with the existing couplings between the various models in \muphy. In this way, the entire hierarchy of kinetic Vlasov simulations on kinetic scales up to MHD on system scales is possible. However, the hierarchy presented in this section is limited to the fluid models integrated in \muphy. Additionally, in this subsection, we adjust the simulations toward more realistic parameters such that a realistic mass ratio $m_i/m_e  = 1836$ is used.

The setup we use is a Harris sheet setup as described in the GEM reconnection challenge in \cite{https://doi.org/10.1029/1999JA900449}.
This perturbed MHD equilibrium requires an accurate kinetic description near the reconnection site to recover the right reconnection physics, but lends itself to an ideal MHD description away from the current sheet.
Beyond the parameters described in the original paper, which are given in terms of MHD variables, the regions not simulated using ideal MHD use a reduced speed of light $c = 20 v_\text{A,0}$, the total current is split between electrons and ions such that $\mathbf{j}_{ez} / \mathbf{j}_{iz} = T_i / T_e$, and the temperature ratio for variable conversion is fixed to the initial ratio $\alpha_T = T_e / T_i$. The AP solver uses $\theta = 1$.

\muphy uses a normalised unit system with the magnetic field and density normalised to characteristic values $B_0$ and $n_0$.
Speeds are given in terms of the Alfvén speed $v_\text{A,0}$, lengths in terms of ion inertial lengths $d_{i, 0}$, times in terms of the inverse ion cyclotron frequency $\omega_{ci}^{-1}$, and electric fields in terms of $v_\text{A,0} B_0$.
Additionally, we use $\mu_0 = 1, k_\text{B} = 1$.
The simulation results in the following sections are given in these units.

\subsection{MHD to Five-Moment Coupling}
\label{sec:MHD-AP-explicit}

The chosen domain size is $L_x = 16 \pi d_\text{i,0}$ and $L_y = 8 \pi d_\text{i,0}$.
Due to the symmetry of the problem, we can simulate just one half of this domain, which we further divide into regions simulated using different models, as shown in the simulation results.
The resolution of this half-domain is set to $256 \times 256$.

\begin{figure}
    \centering
    \includegraphics[width=\textwidth]{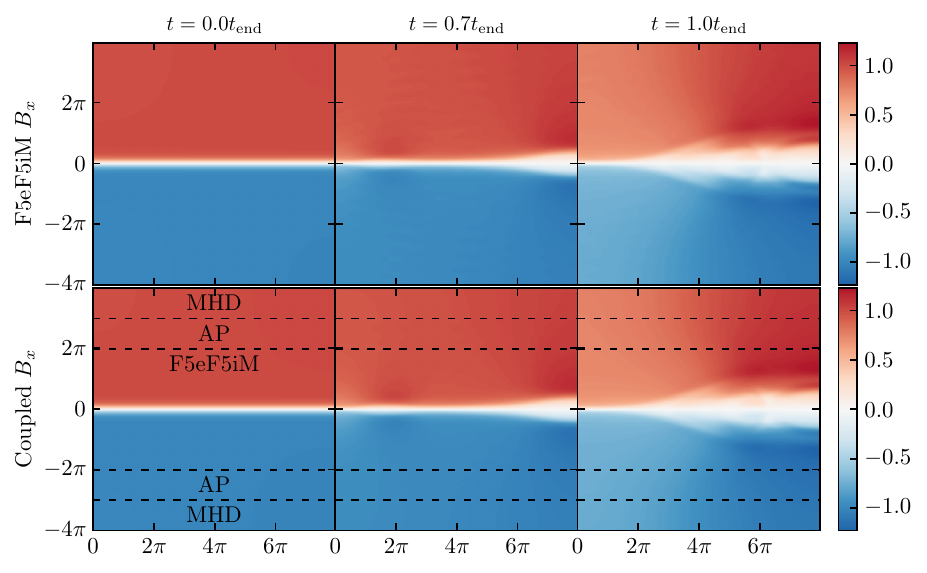}
    \caption{$B_x$ of the coupled and reference simulations of magnetic reconnection.}
    \label{fig:small_reconnection_B_x}
\end{figure}

\begin{figure}
    \centering
    \includegraphics[width=\textwidth]{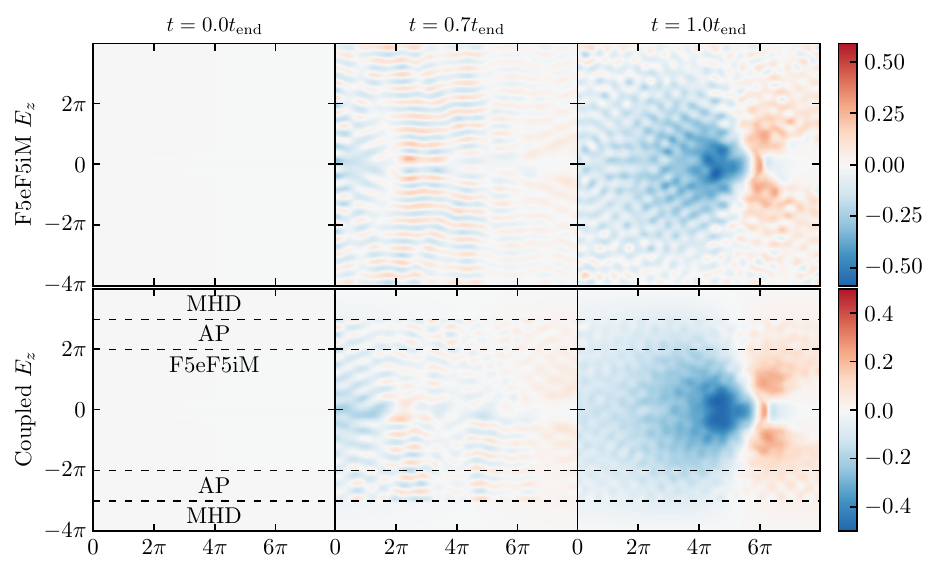}
    \caption{$E_z$ of the coupled and reference simulations of magnetic reconnection.}
    \label{fig:small_reconnection_E_z}
\end{figure}

\begin{figure}
    \centering
    \includegraphics[width=\textwidth]{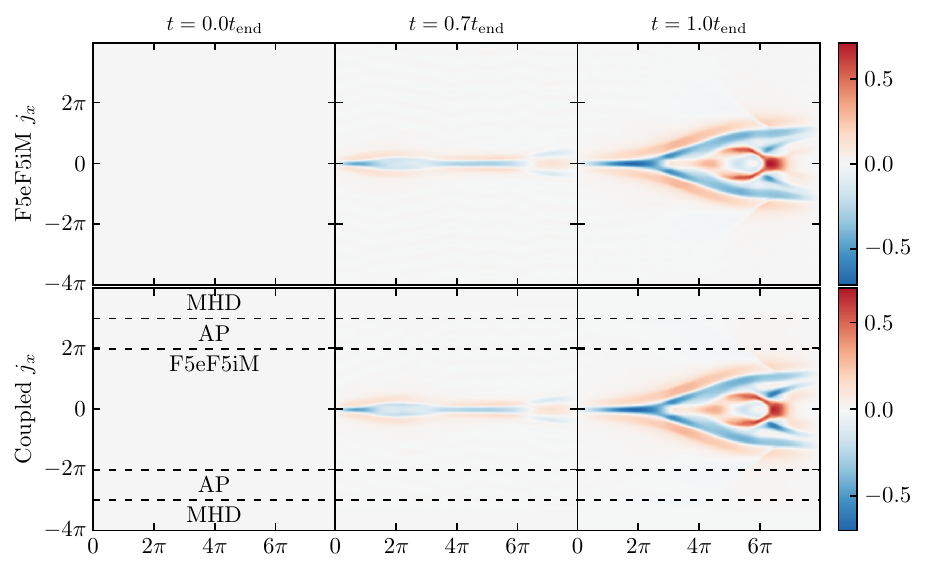}
    \caption{$j_x$ of the coupled and reference simulations of magnetic reconnection.}
    \label{fig:small_reconnection_j_x}
\end{figure}

Fig. \ref{fig:small_reconnection_B_x} - \ref{fig:small_reconnection_j_x} show snapshots of the right half of both the coupled simulation as well as a reference simulation using \texttt{F5eFeiM} throughout the domain, simulated to $t_\text{end} = 40 \Omega_i^{-1}$.
Both the magnetic field (Fig. \ref{fig:small_reconnection_B_x}) and the current (Fig. \ref{fig:small_reconnection_j_x}) match closely between the coupled and reference simulations.
The out-of-plane electric field (Fig. \ref{fig:small_reconnection_E_z}) shows fast oscillations developing as the reconnection starts.
As the Harris sheet setup uses conducting boundary conditions, these waves are reflected at the boundaries of the domain for the reference simulation.
In the coupled simulation, these oscillations are absent from the MHD part of the domain, but they are also reduced within the inner parts of the domain, since they are damped when travelling through the AP regions, significantly reducing the amplitude of waves reflected at the AP-MHD boundary.

\subsection{Testing the full model hierarchy from the 10 moment-two fluid to the MHD model}
\label{sec:MHDto10moment}

\begin{figure}
    \centering
    \includegraphics[width=\textwidth]{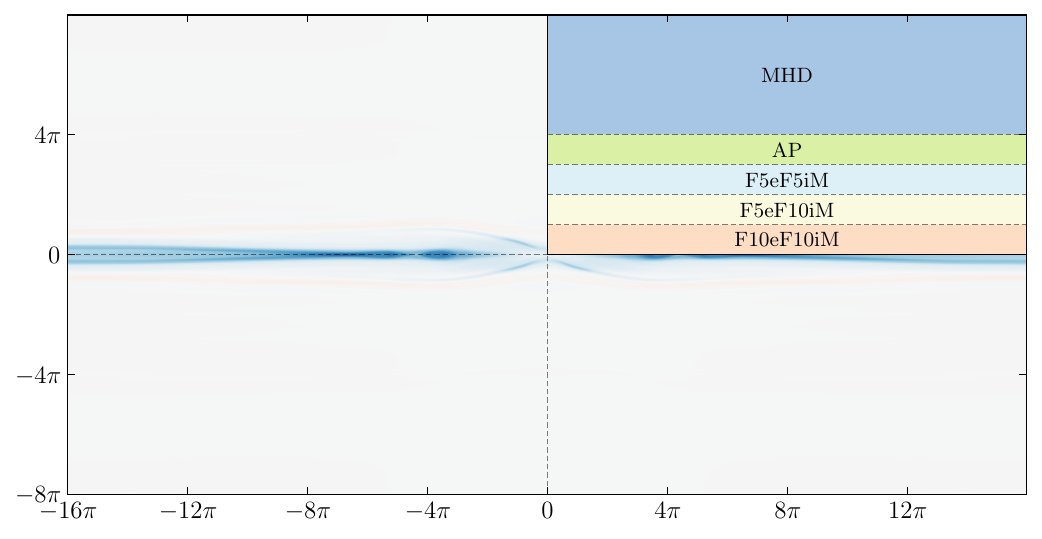}
    \caption{Decomposition of the simulated domain into regions using different schemes.}
    \label{fig:reconnection_decomposition}
\end{figure}

In addition to the change in mass ratio, we double the domain size compared to Sec. \ref{sec:MHD-AP-explicit} to $L_x = 32 \pi d_\text{i,0}$ and $L_y = 16 \pi d_\text{i,0}$, to allow a large portion of the domain to be covered by the MHD solver.
To increase the effective resolution of the simulation, we simulate just one quarter of the domain.
Fig. \ref{fig:reconnection_decomposition} depicts the decomposition of this quarter-domain into regions of different models in the hierarchy.
The resolution of the simulated region is set to $1024 \times 512$.

\begin{figure}
    \centering
    \includegraphics[width=\textwidth]{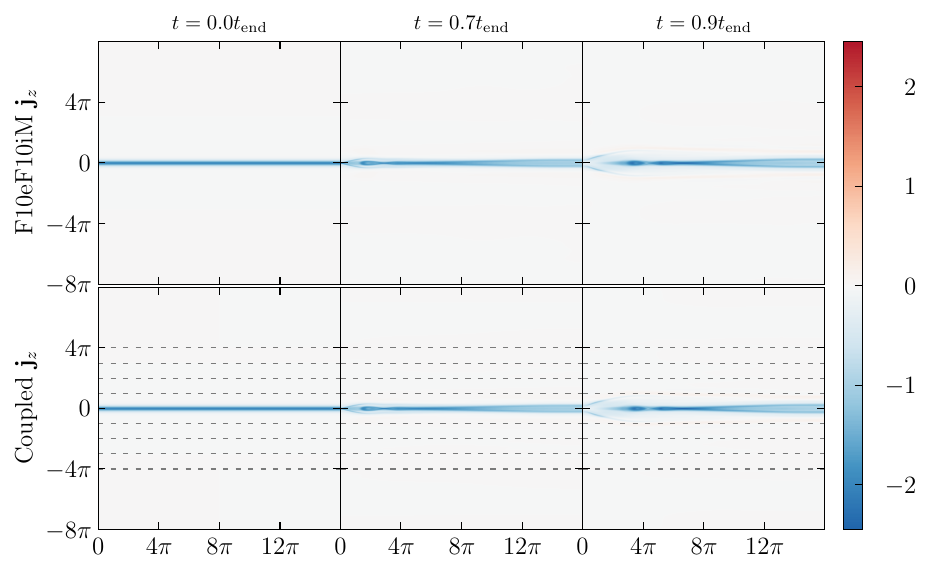}
    \caption{$j_z$ of the coupled and reference simulations of magnetic reconnection.}
    \label{fig:reconnection_j_z}
\end{figure}

\begin{figure}
    \centering
    \includegraphics[width=\textwidth]{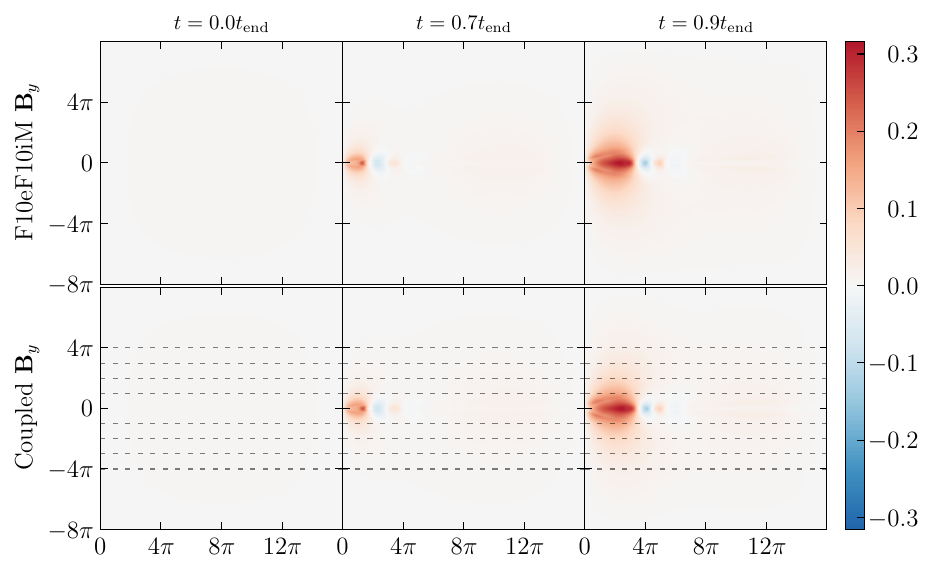}
    \caption{$B_y$ of the coupled and reference simulations of magnetic reconnection.}
    \label{fig:reconnection_B_y}
\end{figure}

\begin{figure}
    \centering
    \includegraphics[width=\textwidth]{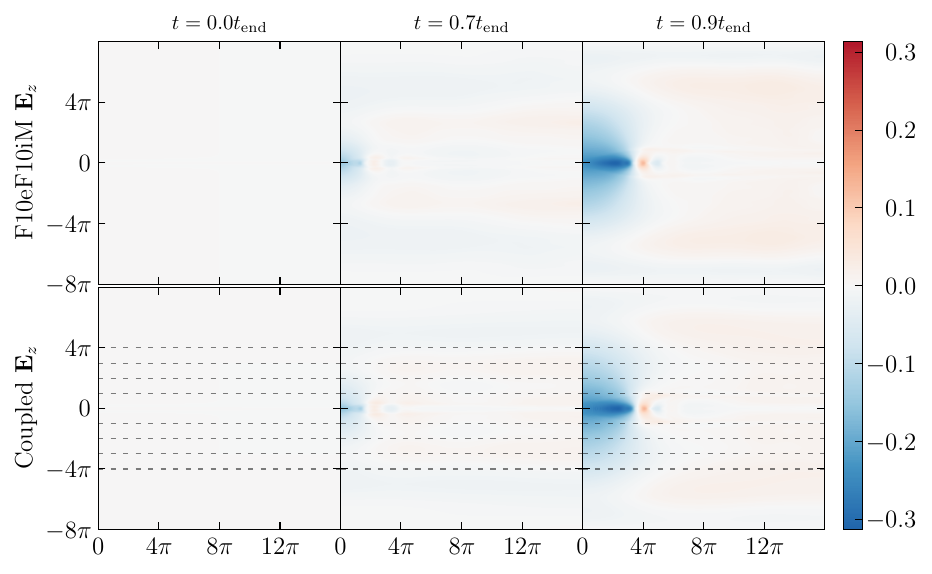}
    \caption{$E_z$ of the coupled and reference simulations of magnetic reconnection.}
    \label{fig:reconnection_E_z}
\end{figure}

Similarly to the previous section, Fig. \ref{fig:reconnection_j_z} - \ref{fig:reconnection_E_z} show snapshots of the coupled simulation as well as a reference simulation using \texttt{F10eF10iM}, simulated again to $t_\text{end} = 40 \Omega_i^{-1}$. The results of the two simulations match closely and none of the transitions between different schemes develop any significant discontinuity or oscillations, also demonstrating that these parts of the domain are well approximated by the different reduced models.

\section{Conclusions and Outlook}
\label{sec:Outlook}

In this work, we have presented a comprehensive multiscale framework for simulating collisionless space plasmas that bridges kinetic-scale physics with large-scale MHD dynamics. Central to our approach is an asymptotic-preserving (AP) formulation of the two-species five-moment equations coupled with Maxwell's equations. We derived a stable numerical scheme that consistently projects fast electromagnetic dynamics onto the slow MHD dynamics, ensuring a seamless and accurate transition to the ideal MHD limit. Building on this, we developed a robust coupling algorithm that interfaces the five-moment solver with an ideal MHD solver, preserving asymptotic consistency and enabling stable variable exchange at the model boundaries.

The proposed methodology was integrated into the \muphy framework and validated through magnetic reconnection simulations. Our results demonstrate that the AP coupling strategy effectively captures localized non-ideal plasma behavior while maintaining the computational efficiency of ideal MHD in large-scale regions. Furthermore, we successfully executed a fully coupled simulation spanning the entire model hierarchy, from fully kinetic and hybrid descriptions down to five-moment and MHD models. This confirms that adaptive model switching, guided by the AP framework, provides a scalable and physically consistent pathway for global plasma simulations.

Looking ahead, the current implementation serves as a functional prototype that demonstrates the viability of the proposed coupling strategy. The next step is to expand upon this architecture by integrating \muphy with more advanced AMR MHD solvers, such as Parthenon/Athena++ \citep{parthenon:2023}, deal.II \citep{dealII:2025}, and Trixi \citep{trixi:2025}. Following the same modular philosophy, this integration will enable dynamic resolution adaptation in large-scale simulations. The underlying MPI infrastructure for a multiple programs, multiple data (MPMD) execution model has already been established to support this, paving the way for efficient, scalable multiscale simulations in space physics.

\section*{Acknowledgments}

We acknowledge funding from the German Science Foundation DFG through the research unit “SNuBIC” (DFG-FOR5409). We gratefully acknowledge the Gauss Centre for Supercomputing e.V. (www.gauss-centre.eu) for funding this project by providing computing time through the John von Neumann Institute for Computing (NIC) on the GCS Supercomputer JUWELS [37] at Jülich Supercomputing Centre (JSC). Computations were conducted on JUWELS/JUWELS-booster and on the DaVinci cluster at TP1 Plasma Research Department.

\bibliographystyle{jpp}
\bibliography{bib.bib}

\end{document}